\documentclass[manuscript]{aastex}

\usepackage{lscape}

\usepackage[usenames,dvips]{color}
\usepackage{graphicx}

\usepackage{color}

\shorttitle{Double-peaked light curve  
of the $\gamma$-ray binary PSR~1259-63/LS~2883 }
\shortauthors{Takata et al.}

\begin{document}


\title{
Modeling high-energy light curves 
 of the PSR~B1259$-$63/LS~2883 binary 
based on 3-D SPH simulations}

 \author{J. \textsc{Takata}\altaffilmark{1},
A.T. \textsc{Okazaki}\altaffilmark{2},
S. \textsc{Nagataki}\altaffilmark{3},
T. \textsc{Naito}\altaffilmark{4},
A. \textsc{Kawachi}\altaffilmark{5},
S.-H. \textsc{Lee}\altaffilmark{3},
M. \textsc{Mori}\altaffilmark{6},
K. \textsc{Hayasaki}\altaffilmark{7},
M. S. \textsc{Yamaguchi}\altaffilmark{8},
   and
S.P. \textsc{Owocki}\altaffilmark{9}
   }
\altaffiltext{1}{Department of Physics, The University of Hong-Kong, Hong Kong}
\email{takata@hku.hk}
\altaffiltext{2}{Faculty of Engineering, Hokkai-Gakuen University, Toyohira-ku,
       Sapporo 062-8605, Japan}
\altaffiltext{3}{Yukawa Institute for Theoretical Physics, Oiwake-cho,
       Kitashirakawa, Sakyo-ku, Kyoto 606-8502, Japan}
\altaffiltext{4}{Faculty of Management Information, Yamanashi Gakuin University,
       Kofu, Yamanashi 400-8575, Japan}
\altaffiltext{5}{Department of Physics, Tokai University, Hiratsuka, Kanagawa
       259-1292, Japan}
\altaffiltext{6}{Department of Physical Sciences, Ritsumeikan University,
 1-1-1 Noji Higashi, Kusatsu, Shiga 525-8577, Japan}
\altaffiltext{7}{Department of Astronomy, Kyoto University, Oiwake-cho, 
Kitashirakawa, Sakyo-ku, Kyoto 606-8502, Japan}
\altaffiltext{8}{Department of Earth and Space Science, 
Graduate School of Science, Osaka University, Toyonaka, Osaka 560-0043, Japan}
\altaffiltext{9}{Bartol Research Institute, University of Delaware, Newark, DE 19716, USA}

\begin{abstract}
 Temporal changes of 
 X-ray to very-high-energy gamma-ray emissions from the pulsar-Be star binary PSR~B1259$-$63/LS~2883  
 are studied based on 3-D SPH simulations 
 of pulsar wind interaction with Be-disk and wind.
 We focus on the periastron passage of the binary and 
 calculate the variation of the synchrotron and inverse-Compton emissions 
 using the simulated shock geometry 
 and pressure distribution of the pulsar wind. The characteristic double-peaked X-ray light curve 
 from observations is reproduced by our simulation under
 a dense Be disk condition 
 (base density $\sim 10^{-9}~\mathrm{g~cm^{-3}}$). 
 We interpret the pre- and post-periastron peaks as being due to a significant increase
 in the conversion efficiency from pulsar spin down power to 
 the shock-accelerated particle energy
 at orbital phases 
 when the pulsar crosses the disk 
 before periastron passage, and when the pulsar wind creates a cavity 
 in the disk gas after periastron passage, respectively.
 On the contrary, in the model TeV light curve, which also shows a double peak feature, 
 the first peak appears around the periastron phase.
 The possible effects of cooling processes on the TeV light curve
 are briefly discussed.
\end{abstract}





\section{Introduction}
 Recent improvement of the techniques of 
  ground-based Cherenkov telescopes has increased 
 the number and variety of TeV gamma-ray objects. 
  Five gamma-ray binaries have been detected so far, with
 all of them known to be high mass X-ray binary systems.    
 Their common emission mechanism has been vastly investigated 
since their discoveries. Each gamma-ray binary is assumed to be composed of 
 a compact object orbiting around a massive star. Among them, 
 PSR~B1259-63/LS~2883 is the only system for which the compact 
 object has been confirmed to be a pulsar.

 PSR~B1259-63 is a 48-ms radio pulsar 
 with a spin down power of $L_{\mathrm{sd}}=8.2\times 10^{35}~\mathrm{erg~s^{-1}}$.
 The spectral type of LS~2883 had been known to be B2V 
\citep{Johnston1994}.
 However,
 a recent precise measurement reports 
 the type to be O9.5V \citep{Negueruela2011}.
 This correction implies a change of the assumed distance to the system 
 from 1.5\,kpc to $2.3\pm 0.4$\,kpc,
 as well as a change of the effective temperature of the star.
 The orbit has a large eccentricity $e$ of 0.87 and a long period 
 $P_{\mathrm{orb}}$ of about 3.4~yr.


Non-pulsed and non-thermal emissions from the binary 
in the radio \citep{Johnston2005, Moldon2011}, 
X-ray \citep{Hirayama1996, Uchiyama2009} 
and TeV energy ranges \citep{Aharonian2005, Aharonian2009} have been
reported, and flare-like GeV emissions 
 were detected around the 2010--2011 periastron passage 
by the Large Area Telescope (LAT) on board
$Fermi$ \citep{Tam2011, Abdo2011}. 
The radio pulse eclipse of about 5 weeks around the periastron 
 suggests that the pulsar goes  to the opposite side 
of the Be-disk plane with respect to the observer,
 while crossing it  plane twice during the course.
 The characteristic double-peaked features 
 observed in the radio and X-ray light curves
 \citep{connors02,chernyakova06}
 can be mainly attributed to the interactions of the pulsar wind and 
 the Be-disk during disk crossings by the pulsar. 
 The peak phases or the peak intensities measured, 
 extensively in the radio band in particular, vary from orbit to orbit, though.
 The observations for 2004 \citep{Aharonian2005} 
and 2007 \citep{Aharonian2009} periastron passage  indicate that  
the TeV light curve  also varies from orbit to orbit. 

The non-thermal emission mechanisms of the system have been 
studied in the framework of leptonic 
\citep[e.g.,][]{Tavani1997,  Khangulyan2007, Sierpowska2008,
 Takata2009, Dubus2010b} and hadronic models (\citealt{Kawachi2004}; 
 see also \citealt{Neronov2007}).
 Unfortunately, Be-disk models which have been adopted so far in this field
 of research are mostly outdated, 
 and are significantly different from the model being 
widely accepted by the Be star research community currently,
 i.e., the viscous decretion disk model
 [\citet{Lee1991}; see also \citet{Porter1999} and \citet{Carciofi2006}]. 
 Since the shock distance and  its geometry depend on the Be-disk model, 
it is important to examine the high-energy emissions with a more 
realistic Be-disk model. For instance, disk models with supersonic outflows, 
as adopted in most previous studies, have irrelevantly large ram pressure, 
and thus give rise to false location of the shock. 


Detailed two-dimensional hydrodynamic simulations have also been performed to study the wind-wind
collision interaction in this system \citep[e.g.,][]{Bogovalov2008,Bogovalov2011}, providing important information on the shock structure.
Since the simulations have been limited to 2-D, however, 
the orbital motion has not been taken into account, which can have 
significant effects near the periastron passage.
Moreover, the presence of the Be-disk can play an important role to the origin
of the high energy emission from this system. In such a case, the behavior of the system
will inevitably manifest in 3-D and be orbital-phase dependent. 

In \citet[][hereafter paper I]{Okazaki2011}, we have 
studied the interaction between the pulsar and the Be star by,
for the first time, carrying out
3-D hydrodynamic simulations using the viscous decretion disk model.
We have found that
for a Be-disk with typical density, the pulsar wind strips off an 
outer part of the Be-disk, truncating the disk at a radius significantly smaller than the pulsar orbit. 
This prohibits the pulsar from passing through the disk around periastron passage, 
which has been assumed in previous studies.
In other words, a Be-disk with typical density is dynamically unimportant 
in this system.

A large H$\alpha$ equivalent width of $-54$\,\AA\ 
\citep{Negueruela2011}, however, suggests that 
the density of the Be-disk of LS~2883 is much higher than typical.
It is, therefore, interesting to see how the interaction 
changes if the Be-disk is much denser and dynamically more important
than those studied in paper~I.
Given the double-peaked light curves of PSR B1259-63, a particularly important
issue is whether a reasonably high Be-disk density can lead to a strong pulsar wind confinement, and hence an enhanced emission, during both disk-plane crossings.

In this paper, which is the second of the series, we study high energy emissions from 
PSR B1259-63/LS 2883 system, based on the results of numerical simulations
for different values of the Be-disk density.
We review our 3-D hydrodynamic simulations in section~\ref{simulation}
 and describe our emission model in section~\ref{model}. 
Based on the model light curves, we propose, in section~\ref{x-ray}, 
a new interpretation of the observed double-peaked feature of the radiation 
(in particular in the X-ray band).
The multi-wavelength emission properties are discussed
in section~\ref{gevtev}. Finally,  we summarize our results in section~\ref{summary}.

\section{Numerical Model for the Hydrodynamic Interaction between the Pulsar and the Be Star}
\label{simulation}

The simulations presented below are performed with a 3-D SPH code. The code is basically identical to
that used by \citet{Okazaki2002} \citep[see also][]{Bate1995}, 
except that the current version is adapted to systems with winds and a decretion disk, such as
PSR B1259-63/LS~2883, and takes into account radiative cooling with 
the cooling function generated by CLOUDY 90.01 for 
an optically thin plasma with solar abundances \citep{Ferland1996}.
Using a variable smoothing length, the SPH equations with a standard
cubic-spline kernel are integrated with an individual time step for
each particle. In our code, the Be-disk, the Be wind, and the pulsar wind are modeled by 
ensembles of gas particles of different particle masses with negligible self-gravity, 
while the Be star and the pulsar are represented by sink particles 
with the appropriate gravitational masses. 
Gas particles which fall within a specified accretion radius are accreted by the sink particle.

To reproduce the Be decretion disk, we inject gas particles just outside of the stellar
equatorial surface at a constant rate. 
In paper~I, we adopted the injection rate of 
$3.5 \times 10^{-9}M_{\odot}\,\mathrm{yr}^{-1}$, which gave rise to a 
typical disk base density of $10^{-11}\,{\rm g~cm}^{-3}$
[detailed model fit of the observed H$\alpha$ profiles typically provides a  density 
between $10^{-12}\,{\rm g~cm}^{-3}$ and several times $10^{-10}\,{\rm g~cm}^{-3}$ 
\citep[e.g.,][]{Silaj2010}]. 
In the current study, however, we compare the resulting radiation spectrum/light curve
for three different injection rates, fixing all the other parameters at values adopted 
in paper~I. Particularly, for the mass and radius of the Be star, we take values typical for a B2V star to ensure
consistency with the models studied in paper~I. The polar axis of the Be star is tilted from 
the binary orbital axis by $45\arcdeg$,
 and the azimuth of tilt, i.e. the azimuthal angle of the Be star's polar axis
from the direction of apastron, is $19\arcdeg$.
With this geometry, the pulsar crosses the equatorial plane of the Be star
at $\tau \sim -10\,\mathrm{d}$ and  $\tau \sim +25\,\mathrm{d}$, 
where $\tau$ is the orbital phase in days relative to the periastron passage.
We choose rates of
$3.5 \times 10^{-9}M_{\odot}\,\mathrm{yr}^{-1}$, 
$3.5 \times 10^{-8}M_{\odot}\,\mathrm{yr}^{-1}$, and
$3.5 \times 10^{-7}M_{\odot}\,\mathrm{yr}^{-1}$, corresponding to 
disk base densities of 
$10^{-11}\,{\rm g~cm}^{-3}$,
$10^{-10}\,{\rm g~cm}^{-3}$, and 
$10^{-9}\,{\rm g~cm}^{-3}$, respectively. 
Remarkably, we find that the highest injection rate is favored
by best reproducing the observed light curve.
We note that taking the base density of $10^{-9}\,{\rm g~cm}^{-3}$ for this system is not unreasonable,
given that LS~2883 showed the H$\alpha$ equivalent width of $-54$\,\AA\ in quiescence 
\citep{Negueruela2011}, which was one of the largest equivalent widths Be stars have ever shown.

The Be wind and pulsar wind 
are turned on in the simulation at a certain time after the Be-disk has fully developed 
in the tidal simulation (for more details, see paper~I). We started the wind simulation at 
$t=11.44\,P_{\rm orb}$, 74 days prior to periastron passage.
For simplicity, we assume that the winds coast without any net external force, 
assuming in effect that gravitational forces are 
either negligible (i.e. for the pulsar wind) or are cancelled by radiative driving terms (i.e. for the Be wind).
The relativistic pulsar wind is emulated by a non-relativistic flow with 
a velocity of $10^{4}\,{\rm km s}^{-1}$ and an adjusted mass-loss rate 
so as to provide the same momentum flux as a relativistic flow 
with the same assumed energy.
We assume that all the spin down energy 
$L_{\rm sd}=8.2 \times 10^{35}\,{\rm erg~s}^{-1}$ goes to 
the kinetic energy of a spherically symmetric pulsar wind. 
We also assume the Be wind to be spherically symmetric with 
a mass loss rate of $10^{-8}\,M_{\odot}\,{\rm yr}^{-1}$.

 It is noted that the  unshocked  pulsar wind  
is emulated by a non-relativistic flow with a momentum flux ($L_{\rm sd}/c$), 
where $c$ is the speed of light.  The global structure of 
pulsar wind under the influence of the Be-wind/disk 
should  be  reasonably represented by this treatment. 
This is because it depends primarily on the 
momentum flux ratio and not on whether the pulsar wind is modeled 
as relativistic (see paper I for details). This will 
 justify the application  since we estimate the light curve
and spectrum by integrating all the contributions from each emission region.
 The resulting light curve and spectrum should not depend on the detailed local
 structure of the shocked pulsar winds. Rather, they should depend mainly on 
the global structure. Thus we conclude
 the resulting  light curve and spectrum 
 in this study are robust in spite of our non-relativistic treatment. 
Of course, as stated in Bogovalov et al. (2008, 2011), the relativistic effects 
can affect the detailed structure of unshocked/shocked pulsar wind regions. 
Thus it is 
our future plan to extend our code to the  relativistic regine and examine its 
influence. The possibility of relativistic Doppler boost effect is discussed 
in section 4.1. 

\section{Emission Model for Shocked Pulsar Wind}
\label{model}
We calculate the emission at each orbital phase
using data from the 3-D hydrodynamic simulations. 
First, the simulation volume is divided into uniform grids,
and at each grid point the pulsar wind pressure is calculated by counting 
only the contribution from the pulsar wind particles.
Then, from the 3-D distribution of the pulsar wind pressure,
synchrotron and inverse-Compton emissions are locally
evaluated by using an assumption on the local magnetic 
fields and the calculation scheme
described in the following subsections.
In these calculations, we implicitly assume
 that particles are accelerated through the 1st-order Fermi
mechanism at the pulsar wind termination shock, and have a power-law
energy distribution.  We anticipate that the motion of each SPH 
particle, which represents an ensemble of electrons and positrons, corresponds 
to the bulk motion of the pulsar wind particles represented by the 
SPH particle.   
In the down stream region, we assume that 
the  motion of the electrons and positrons represented by each SPH particle
 is randomized and the energy distribution is described by the power law 
function (section~\ref{dis}).

The total emissions are obtained by integrating the local
 emissions over the whole simulation volume. We adopt $100^{3}$ 
uniform grids and have confirmed that 
the result does not change even if grids with higher resolution are used.
Although the integration is taken  over the whole simulation volume, 
virtually all the contribution to
  the high energy emission arises from the shocked pulsar wind region,
  because the unshocked (upstream) region of the pulsar wind is too
  cold to make any significant contribution to the emission. 
It has been suggested that the relativistic bulk motion of the upstream  
flow may produce a considerable high energy  emissions
 via the inverse-Compton process. 
In this study, however, because 
 our  simulations are done in the non-relativistic limit, we do not 
investigate this point.  
It has been suggested that the radio emission can come from 
 a larger volume than emissions in other wavelengths 
 \citep{Moldon2011}.
 This volume is roughly $\sim$ 100\,AU,  which is about one order of magnitude larger 
 than the size of the present simulation volume ($\sim 2a\sim 1.5\times 10^{14}$~cm).
 Estimation of emission in the radio band is hence omitted here and postponed to future studies.

\subsection{Magnetic Fields}

Our simulations are performed in the hydrodynamic limit and 
 there are theoretical uncertainties involved in determining the magnetic field 
 of the pulsar wind.
 The magnetic field in the shock down-stream 
 is obtained using the Rankine-Hugoniot relations at the shock surface 
 and the adiabatic expansion-law of ideal MHD flow 
 as in the cases of steady nebulae 
(e.g. Kennel and Coroniti 1984;  Tavani and Arons 1997;
 Nagataki 2004).
 The flow in our simulations, however, is too dynamic for the above
 calculation scheme to be applied to. Therefore, 
 we adopt another approach frequently used in modeling 
  gamma-ray bursts (e.g. Sari et al. 1998; Xu et al. 2011) 
as follows.  On the pulsar side, the total pulsar wind pressure $P_{\rm tot}$ 
at each  grid is associated with the local magnetic field $B$  as 
\begin{equation}
B=\sqrt{\eta \times 8\pi P_{\rm tot}}, 
\label{magnetic}
\end{equation}
where $\eta$ is fixed to 0.1, which gives a X-ray flux consistent 
with observations of PSR~B1259-63/LS 2883 system.

\subsection{Scheme of Calculation}

\subsubsection{Energy distribution of accelerated particles}
\label{dis}
The synchrotron  and inverse-Compton processes are calculated in the 
  simulation grids where the   pulsar wind pressure 
$P_\mathrm{tot}$ is non-zero  and  the contribution of the 
wind particles is $P_\mathrm{g}\equiv (1-\eta)P_\mathrm{tot}$.
  We assume that  the number of  shocked 
 particles per Lorentz factor per volume at each grid point
 is given by a single power law function of
 \begin{equation}
f(\Gamma)=\frac{K}{4\pi}\Gamma^{-p}~~(\Gamma_\mathrm{min}\le \Gamma\le \Gamma_\mathrm{max}),
\label{func}
\end{equation}
where $\Gamma$ is the Lorentz factor of the particles.
Note that the accelerated particles are strictly relativistic in our model, such that their momenta are simply $m_e c \Gamma$, where $m_e$ is
 the electron rest mass. 
We will therefore write down their distribution in $\Gamma$ instead of momentum for convenience. 

The minimum Lorentz factor $\Gamma_\mathrm{min}$ is similar to
 the Lorentz factor of the bulk motion of the
 unshocked flow.  At  the limit of small magnetization parameter 
($\sigma\ll 1$),  the latter is given by  $\sim \sigma_L\Gamma_L$,
 where $\sigma_\mathrm{L}$ and $\Gamma_\mathrm{L}$
 are the magnetization parameter and the bulk Lorentz factor 
at the light cylinder radius of the pulsar.
 With  $\sigma_\mathrm{L} \sim 10^{3}$ and $\Gamma_\mathrm{L} \sim 10^{2-3}$, 
 $\Gamma_\mathrm{min}$ of 5$\times 10^{5}$ is obtained.
The maximum Lorentz factor 
 $\Gamma_\mathrm{max}$ is determined as 
  $\Gamma_\mathrm{max}=\mathrm{min}(\Gamma_\mathrm{s},\Gamma_\mathrm{c})$;
  $\Gamma_s$  is the Lorentz factor
 at which the acceleration timescale $t_a\sim \Gamma m_ec/eB$ with $e$ being 
electron charge is equal to  the synchrotron loss
 timescale $t_s\sim 9m_e^3c^5/4e^4B^2\Gamma$,
  whereas  $\Gamma_c$ is defined as 
the Lorentz factor at which 
 the acceleration timescale $t_a$ is equal to the dynamical timescale 
 of the shocked pulsar wind  $t_c$.
  Here, $t_c$ is deduced as $\sim h_p/(c/\sqrt{3})$ with 
 the scale height $h_p$ of the gas pressure 
 estimated from the simulation. The typical  $\Gamma_\mathrm{max}$
 in this work is of order 10$^8$.

 The energy density $\epsilon_p$ is related to the 
 gas pressure $P_g$ as $\epsilon_p=3P_g$. Using the 
 condition $\epsilon_p=m_ec^2\int\int \Gamma f(\Gamma)d\Gamma d\Omega$, we 
 calculate the normalization factor $K$ as
\begin{equation}
K=\frac{3P_g}{m_ec^2}\left \{
\begin{array}{@{\,}ll}
\frac{2-p}{\Gamma_{\rm max}^{2-p}-\Gamma_{\rm min}^{2-p}}
& \mathrm{for}~~p\ne 0 \\
\left[\mathrm{ln}(\Gamma_{\rm max}/\Gamma_{\rm min})\right]^{-1}& \mathrm{for}~~p=2
\end{array}
\right .
\label{norm}
\end{equation}

It is well known that the standard  1st-order Fermi mechanism 
in the test-particle limit gives $p=(r+2)/(r-1)$, 
where $r$ is the compression ratio at the shock wave.
Assuming a strong, non-relativistic shock, the compression ratio 
$r=4$ is derived  from the Rankine-Hugoniot relation, 
so that the index$p=2$. From previous studies, in general we can expect that
$p \geq 2$ for weak, non-relativistic shocks, and 
$1.0 \leq p \leq 2.2$ for strong, relativistic shocks, 
respectively (see, e.g. Longair 1994 and references therein). 
Limited by the spatial resolution, nevertheless,
it is very difficult to determine the shock conditions in the emission region directly from the hydro simulation.  
Therefore, unless mentioned otherwise, we will adopt $p=2$ as our canonical value for the results in this paper. 
In addition,  cases for $p=2 \pm 0.5$ will also be investigated to 
illustrate the possible effects of the uncertainty in $p$ on our 
results, in particular the multi-wavelength emission spectra.

 \subsubsection{Formulae}

 The synchrotron power per unit energy emitted by each electron is 
calculated according to Rybicki \& Lightman (1979) as 
\begin{equation}
P_{\rm syn}(E)=\frac{\sqrt{3}e^2B\sin\theta_p}{hm_ec^2}F_{\rm sy}
\left(\frac{E}{E_{\rm syn}}\right),
\end{equation}
where $E_{\rm syn}=3he\Gamma^2B\sin\theta_p/4\pi m_ec$ 
is the typical photon energy
 and $F_{\rm sy}(x)=x\int_x^{\infty}K_{5/3}(y)dy$, where $K_{5/3}$ is the modified Bessel function of order 5/3. For the pitch angle $\theta_p$, we use the averaged value corresponding to $\sin^2\theta_P=2/3$. The power per unit energy and per unit solid angle of the inverse-Compton process is described by \citep{Bege1987}

\begin{equation}
\frac{dP_{\rm IC}}{d\Omega}={\mathcal D}^2\int_0^{\theta_c}(1-\beta\cos\theta_0)I_b/h\frac{d\sigma'_{KN}}{d\Omega'}d\Omega_0,
\end{equation} 
where  $d\sigma'/d\Omega'$ is the differential Klein-Nishina cross section, 
 $\beta=\sqrt{\Gamma^2-1}/\Gamma$, 
$\mathcal{D}=\Gamma^{-1}(1-\beta\cos\theta_1)^{-1}$ 
with $\theta_1$ and $\theta_0$ being the angles between the direction of the particle motion 
and the propagating direction of the scattered photon and the background 
photons, respectively,  $h$ is the Planck constant, 
 $I_b$ is the background photon field 
and  $\theta_c$ expresses the angular size of star as seen from the emission 
point. For the target stellar photons, a soft photon field 
 with an effective temperature of 30,000~K is taken, 
 and the Be-disk emission is omitted for simplicity.

For the Earth viewing angle,  we assume that 
 the inclination angle of the orbital plane with respect to the sky 
 is $i\sim 23^{\circ}$ and the true anomaly of the direction of Earth is about 
 $\phi\sim 130^{\circ}$ (Johnston et al. 1996; Negueruela et al. 2011). 
The model spectrum of the emission from the shocked wind measured at 
Earth is calculated as 
\begin{equation}
F_E(E)\sim\frac{e^{-\tau_{\gamma\gamma}}}{D^2}\Sigma_i
\left [ \delta V_i \int_{\Gamma_{\rm min}}^{\Gamma_{\rm max}} f(\Gamma)\left(
P_{\rm syn}+\int\frac{dP_{\rm IC}}{d\Omega}d\Omega \right)d\Gamma\right],
\end{equation}
where $\Sigma_i$ expresses the summation of the each grid and $\delta V_i$ is 
the volume of the each grid. 
The optical depth, $\tau_{\gamma\gamma}$ for the pair-creation 
process between the gamma-rays 
 emitted by the wind particles and the stellar photons is expressed
 by 
\begin{equation}
\tau_{\gamma\gamma}=\int_0^{\infty}d\ell\int_{E_{\rm c}}^{\infty}dE_s
\sigma_{\gamma\gamma}dN_s/dE_s,
\end{equation}
where $\ell$ is the propagating distance of the $\gamma$-ray from the 
emitted point, 
 $dN_s/dE_s$ is the distribution of the number  density 
of the stellar soft photon   and 
\begin{equation}
\sigma_{\gamma\gamma}(E_{\gamma},E_s)=\frac{3}{16}\sigma_{\rm T}(1-v^2)
\left[(3-v^4)\mathrm{ln}\frac{1+v}{1-v}-2v(2-v^2)\right],
\end{equation}
where $\sigma_T$ is the Thomson cross section,  
$v(E_{\gamma},E_s)=\sqrt{1-E_{\rm c}/E_\gamma}$
and $E_{\rm c}=2(m_ec^2)^2/[(1-\cos\theta_{\gamma\gamma})E_s]$
 with $\theta_{\gamma\gamma}$ being the collision angle.
  We use the distance $D=2.5$~kpc in this work.

\section{Results and Discussion}
\label{result}
In this section, we first summarize the results of the hydrodynamic interaction
between the pulsar wind and the circumstellar material of the Be star.
We then present the multi-wavelength light curves and an time-averaged spectrum 
obtained using the simulation data.
We will also briefly discuss the cooling processes and comment on our non-relativistic emulation of the relativistic pulsar wind.

\subsection{Hydrodynamic Interaction between the Pulsar and the Be star}
\label{interaction}

\begin{figure}
\begin{center}
\includegraphics[height=10cm]{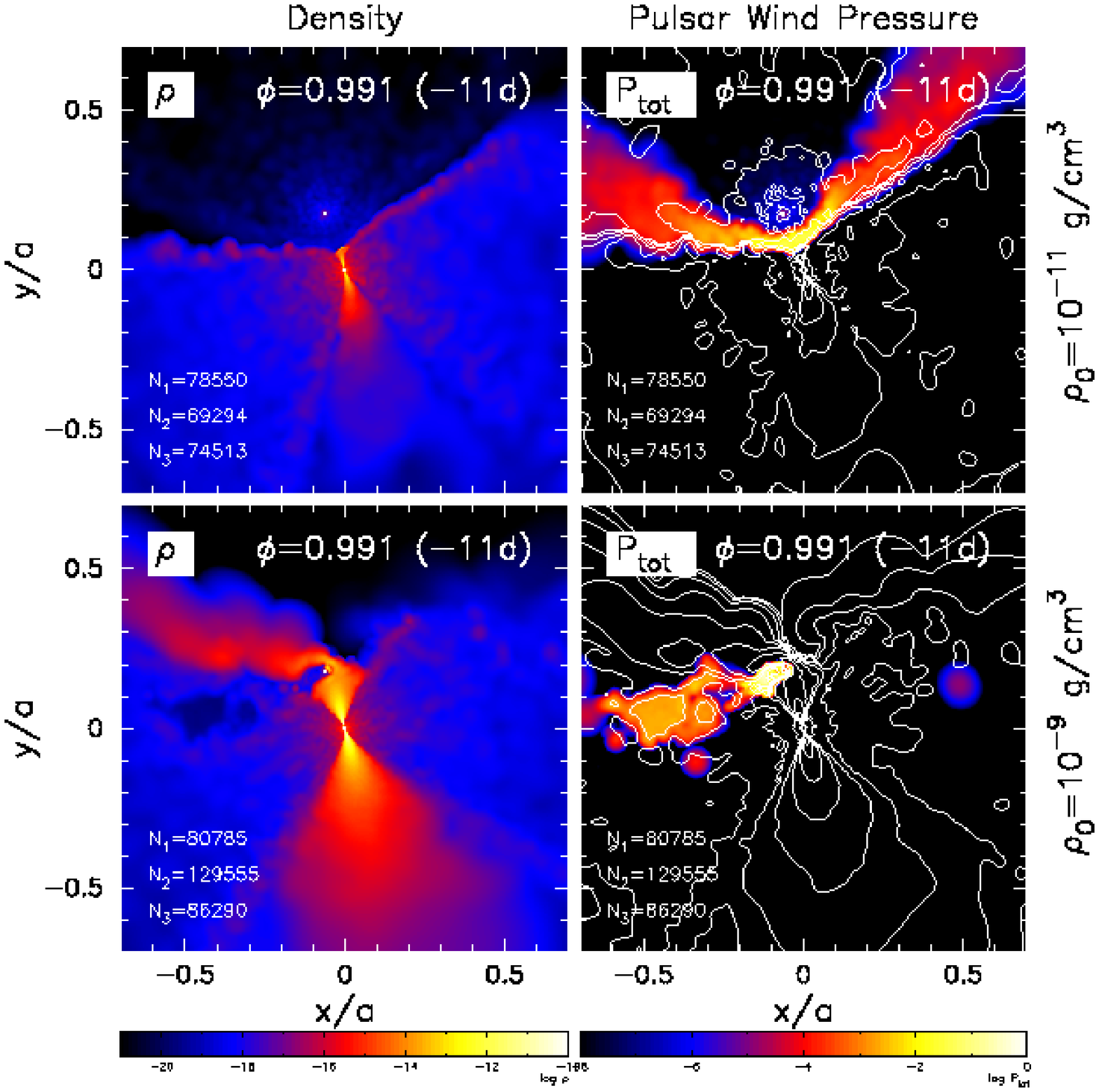}
\caption{Snapshots taken from two SPH simulations for different 
  base densities of the Be-disk: $\rho_0=10^{-11}\,\mathrm{g~cm}^{-3}$
 (upper panels)  and $\rho_0=10^{-9}\,\mathrm{g~cm}^{-3}$ (lower panels).
  The left and right panels show the volume density and 
the pulsar wind pressure,   respectively, 
at $\tau=-11\,\mathrm{d}$ (11 days prior to the periastron).
  In the left panels, the locations of the Be star (the central white dot) 
  and the pulsar (the white dot above the Be star) are also shown, while
  in the right  panel, the volume density is  denoted by contours.
  The density  varies by one order of magnitude per contour.
  In each panel, $N_1$, $N_2$, and $N_3$ annotated at the lower left corner are
  the numbers of particles in the Be wind, the pulsar wind, and 
  the Be-disk, respectively.}
\label{gamma-t-11d}
\end{center}
\end{figure}

\begin{figure}
\begin{center}
\includegraphics[height=10cm]{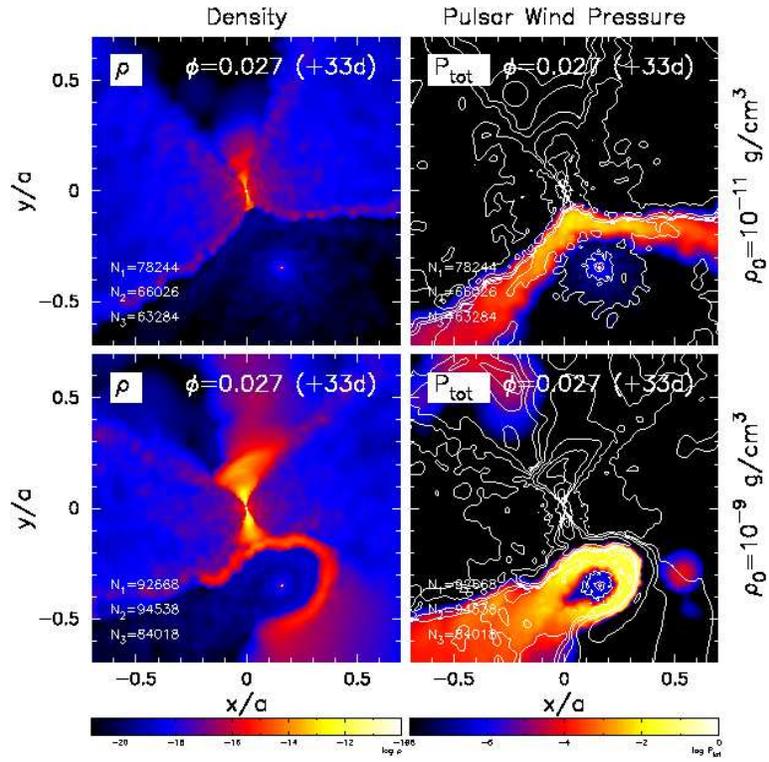}
\caption{Same as Figure~\ref{gamma-t-11d}, but at $\tau=+33\,\mathrm{d}$
 (33 days after the periastron).}
\label{gamma-t+33d}
\end{center}
\end{figure}

As mentioned in section~\ref{simulation}, we have run 3-D SPH simulations
of hydrodynamic interaction in PSR~B1259-63/LS~2883 with the base density
of the Be decretion disk in the range of $10^{-9}-10^{-11}\,\mathrm{g~cm}^{-3}$.
Figures~\ref{gamma-t-11d} and \ref{gamma-t+33d} show snapshots 
of the interaction between the pulsar wind and the circumstellar material of the Be star in the binary orbital plane at two different 
epochs, 11 days prior to periastron passage ($\tau = -11~\mathrm{d}$) 
and 33 days after it ($\tau = +33~\mathrm{d}$), respectively.
Each figure compares the shock structure in two SPH simulations with 
different disk base densities, $\rho_0=10^{-11}\,\mathrm{g~cm}^{-3}$ (upper panels)
and $\rho_0=10^{-9}\,\mathrm{g~cm}^{-3}$ (lower panels):  left and  right 
 panels show the distribution of the volume density and  
the pulsar wind pressure, respectively.  To clarify the location where 
the pulsar wind is terminated, the right panel also shows the distribution 
of the volume density in contours. These figures highlight the effect of 
the Be-disk density on the geometry of 
the interaction surface.
For $\rho_0=10^{-11}\,\mathrm{g~cm}^{-3}$, the pulsar wind easily strips off 
an outer part of the Be-disk, truncating the disk at a radius significantly
 smaller than the pulsar orbit. As a result, the interaction 
surface is open and covers only a small solid angle around 
the pulsar (Figures~\ref{gamma-t-11d} and \ref{gamma-t+33d}, upper panels), 
implying that only a small fraction of the
pulsar wind energy is available for particle acceleration.

In contrast, for $\rho_0=10^{-9}\,\mathrm{g~cm}^{-3}$, the Be-disk
has a large enough inertia not to be pushed away so easily by the pulsar wind.
Thus, as the pulsar approaches the Be-disk before periastron passage,
 the distance between the interaction surface and the pulsar rapidly 
decreases and finally becomes smaller than the scale-height of the disk. 
The pulsar then penetrates the Be-disk, opening a small cavity around it. 
The disk gas surrounding the pulsar terminates the pulsar wind over a 
large solid angle, converting a large fraction of the bulk pulsar wind 
energy into the energy of shocked particles. After periastron, 
the pulsar approaches the Be-disk again, now moving away from the Be star. 
Since the pulsar wind pushes the disk in the direction along which the disk 
density decreases, it can move the disk gas more easily than it could
 before periastron. As a result, a slowly expanding shell is formed around 
the pulsar, which terminates the pulsar wind again over a large solid angle, 
efficiently converting the pulsar wind energy into the energy of shocked 
particles (Figure~\ref{gamma-t+33d}, lower panels). The details of these  
hydrodynamic simulations will be discussed in a subsequent 
paper (Okazaki et al. 2012, in prep)

\subsection{X-ray Light Curves}
\label{x-ray}

\begin{figure}
\begin{center}
\includegraphics{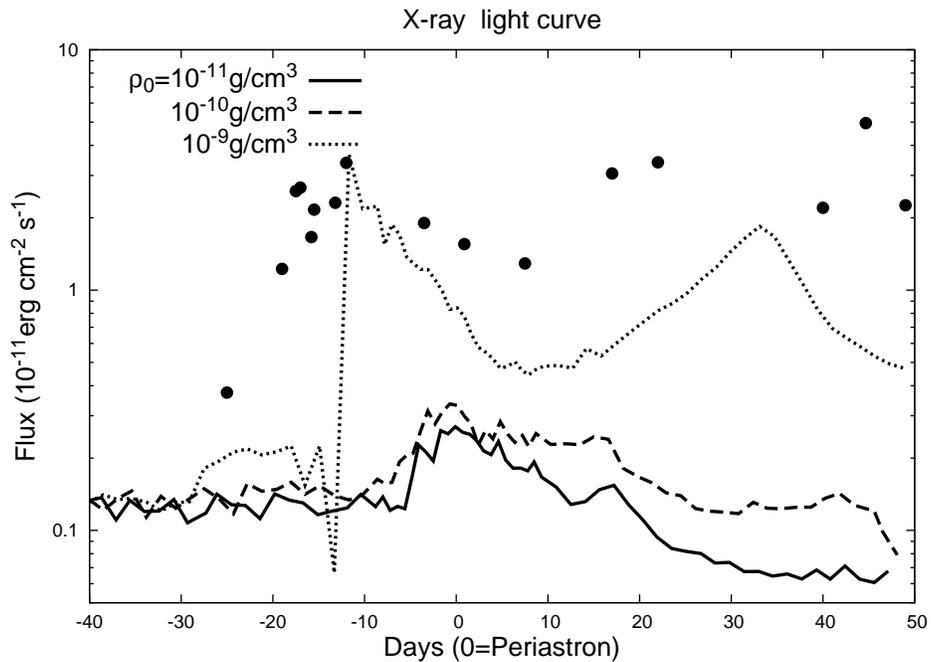}
\caption{The light curves in the 1-10~keV energy band. The solid, dashed and 
dotted lines are the results for base densities of $\rho_0=10^{-11}~\mathrm{g cm^{-3}}$, $10^{-10}~\mathrm{g cm^{-3}}$ and $10^{-9}~\mathrm{g cm^{-3}}$, respectively. We assumed a power index $p=2$ for the 
accelerated particles. Overlaid in the plot are data points from 
Neronov \& Chernyakova (2007).}
\label{light}
\end{center}
\end{figure}

 Figure~\ref{light} presents the calculated light curves of 
 the X-ray flux (1-10~keV energy bands) together with the observed fluxes 
taken from Neronov\& Chernyakova (2007)
 as a function of days relative to the periastron passage, $\tau$.
 The solid, dashed  and dotted lines are 
 the results for disk base densities 
 of $\rho_0=10^{-11}~\mathrm{g~cm^{-3}}$,
 $10^{-10}~\mathrm{g~cm^{-3}}$ and $10^{-9}~\mathrm{g~cm^{-3}}$, respectively, 
 with the typical value of the power index $p=$2 for the accelerated particles.

 During the pre-periastron period, up to  $\tau=-11\,\mathrm{d}$, 
 the flux does not depend on the disk base density. 
 This is because the pulsar wind interacts mainly with the 
  stellar wind, for which 
 we assume an identical  mass loss rate
 of $10^{-8}\,M_{\odot}\,{\rm yr}^{-1}$.
 The epochs during which the pulsar crosses the Be-disk plane 
 are estimated by the simulation to be 
 about 11 days prior to the periastron passage ($\tau \sim -11\,\mathrm{d}$) 
 and 25 days after it ($\tau \sim +25\,\mathrm{d}$). 
 The interaction between the pulsar wind 
 and Be-disk is expected to be the strongest during the disk crossings.
 
Remarkably, however, 
the X-ray flux for cases of disks with typical base densities $\rho_0 = 10^{-10} - 10^{-11}$ $\mathrm{g~cm^{-3}}$
does not peak at the timing of disk crossings,
but shows a maximum intensity at the periastron. On the other hand,
the X-ray flux for the highest density case 
($\rho_0=10^{-9}~\mathrm{g~cm^{-3}}$) 
increases distinctively around the phases of the disk crossings, 
resulting in a double-peaked structure.

\begin{figure}
\begin{center}
\includegraphics{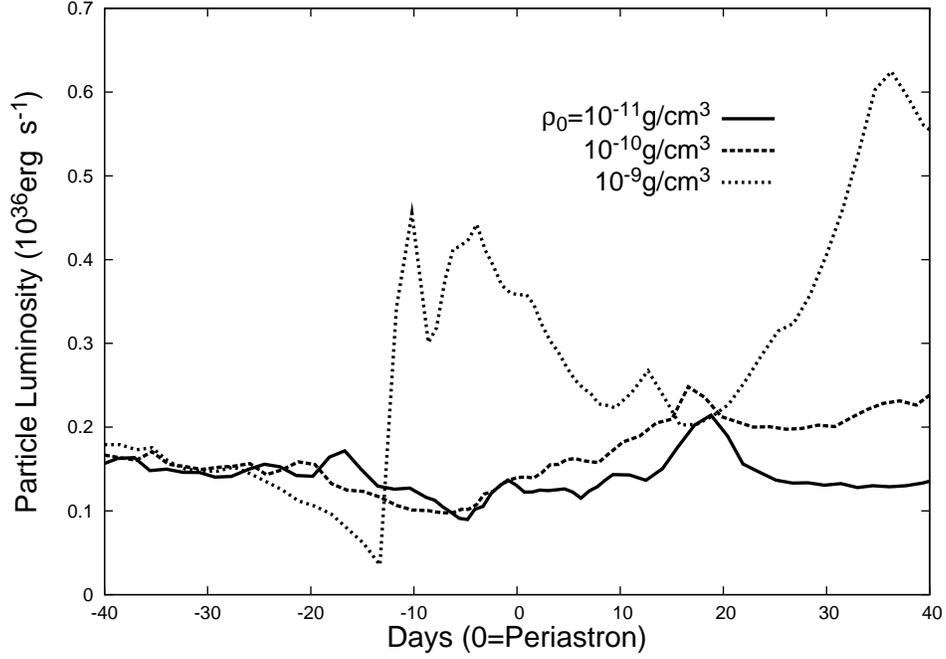}
\caption{ 
Energy budget for the radiation vs. the orbital phase. 
The vertical line corresponds to the physical quantity $\Sigma_i (c/3)P_{\mathrm{tot},i} \delta V_i^{2/3}$, where $\delta V_i$ is the volume of the grids.
The solid, dashed and dotted lines are results for base densities of $\rho_0=10^{-11}~\mathrm{g cm^
{-3}}$, $10^{-10}~\mathrm{g cm^{-3}}$ and $10^{-9}~\mathrm{g cm^{-3}}$, 
respectively.}
\label{ptot}
\end{center}
\end{figure}

 The dependence of the light curves on the Be-disk density 
 reflects that of the shock geometry and resultant 
 conversion efficiency from the spin down power to the internal energy of 
the shocked pulsar wind (Figure~\ref{ptot}).
 For a typical disk density, we see that the pulsar wind can easily 
truncate    the Be-disk at a radius smaller than the pulsar's orbit. 
Therefore, the solid angle as measured from the pulsar, 
 over  which the pulsar wind is stopped by the Be-disk, is small. In this case, the intensity of synchrotron emission is highest
at the periastron because the magnetic fields of the shocked pulsar wind
are highest there (see equation (1)).

 The pulsar wind cannot dismiss the disk with 
 the higher density ($\rho_0=10^{-9}~\mathrm{g~cm^{-3}}$), 
 i.e., with a large inertia. 
 The shock is pushed back toward the pulsar
 in the first disk plane crossing ($\tau \sim -11\,\mathrm{d}$).
 After the periastron, the pulsar approaches the Be-disk again, 
 but now moving away from the Be star.
  Since  the pulsar wind pushes the disk in the direction
 along which the disk density decreases,
 it displaces the disk gas more easily than it did in the 
 pre-periastron crossing.
 As a result, a slowly expanding shell is formed around the pulsar, 
 which terminates the pulsar wind over a larger solid angle
 ($\tau \sim +33\,\mathrm{d}$).
 The particles at the shock obtain energy from
 the bulk kinetic energy of the unshocked pulsar wind, which is in turn provided by
 the pulsar spin down power.
 At these disk-plane crossing phases,  
 the conversion efficiency from the pulsar spin down power to 
 the shock-accelerated particle energy
 drastically increases, which leads to the increase in the X-ray flux, 
 most notably in the post-periastron crossing phase (Figure~\ref{ptot}).

We would like to stress that no fine-tuning has been done in our study 
for the sake of reproducing  the observed X-ray light curve. Thus it is
 remarkable that the calculated flux level and double-peak phases
 for the case of $\rho_0=10^{-9}~\mathrm{g~cm^{-3}}$ and $\eta=0.1$ in equation~(\ref{magnetic}) turn out to be very similar to those from observations,
as shown in Figure~\ref{light}. Our model sheds new light on the disk density 
as a probe of the high energy emission mechanisms. Interestingly, the favored density in this
study is higher than the typical one for Be-disks.

Our model possesses a few profound features
which distinguish it from previous studies. 
First, the present simulations predict that the conversion efficiency
varies with the orbital phase, while the previous studies have
assumed a constant conversion efficiency over the whole orbit
\citep[e.g.,][]{Tavani1997, Takata2009}.
Second, in the present model,
the conversion efficiency and therefore the X-ray flux
acquire the maximum values when the pulsar wind interacts
 with the Be-disk. In \citet{Tavani1997}, on the other hand,
 the inverse-Compton cooling process which
 dominates over other cooling processes 
 is crucial to reproduce the observed decrease of the flux
 near the periastron.
\citet{Takata2009} and \citet{Kong2011} invoked the model that the pulsar
wind parameters at the shock (e.g. the $\sigma$ parameter) vary throughout the
orbital phase. 

\subsection{GeV/TeV Light Curves}
\label{gevtev}
\begin{figure}
\begin{center}
\includegraphics{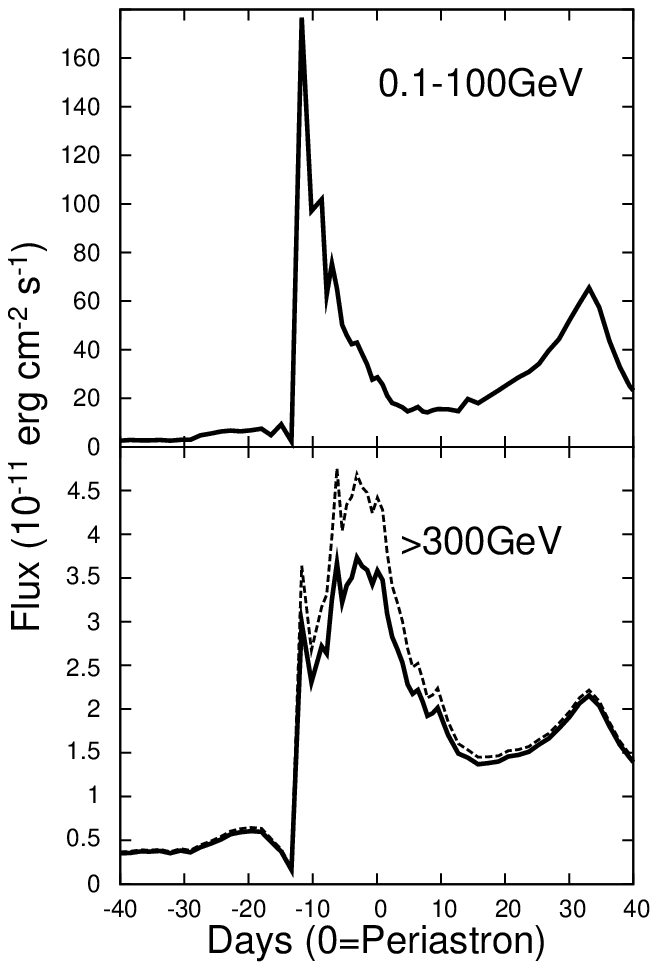}
\caption{The model light curves for different energy bands 
for $\rho_0=10^{-9}~\mathrm{g~cm^{-3}}$. The upper 
and lower panels show the light curves in the 0.1-100~GeV  and $>$300~GeV
 energy bands, respectively. In the lower panel, the dotted line 
represents  the light curve without accounting for 
the effect of photon-photon absorption. The results are for power index 
$p=2$ for the accelerated particles.}
\label{depen}
\end{center}
\end{figure}

 Figure~\ref{depen} shows the calculated light curves for 
0.1-100~GeV    and TeV ($>$300GeV) energy bands.
 We find that the gamma-ray light curves show the double-peaked structures
 as in their X-ray counterpart.
 For 0.1-100~GeV light curves, the peaks 
 align with those in the X-ray bands, because both  X-ray  and 100~MeV-1~GeV 
  emissions are produced by the synchrotron process. 
  In the recent results from $Fermi$
 \citep{Tam2011, Abdo2011}, however, it is likely that 
the phases of the peaks in the X-ray and the GeV bands 
 do not align with each other.   Furthermore, $Fermi$ has not detected 
a strong and sharp peak before the periastron, which is 
 seen in the present model. Hence, in the present framework, 
 there are some discrepancies between the properties of the calculated 
and observed GeV light curves.
  We would like to note that 
the 0.1-1~GeV flux  particularly depend on  the magnetic field, 
 because the synchrotron spectrum from the shocked particles
 has a cut-off around 1-200~MeV.  
In the present model, if the maximum Lorentz factor is determined 
by the balance between the 
synchrotron cooling timescale and the acceleration timescale, 
 the maximum Lorentz factor 
of the particles is expressed as
 $\Gamma_{\mathrm{s}}\sim (9m_e^2c^4/4e^3B)^{1/2}$, which indicates 
the synchrotron photon energy of $E_{\mathrm{syn}}=
3\Gamma_{\mathrm{s}}^2eB\sin\theta_p/(4\pi m_ec)
\sim 3^{5/2}m_ec^3h/(2^{7/2}\pi e^2)\sim 200$~MeV.
For a lower magnetic field, the maximum Lorentz factor may be 
 determined by the balance between the acceleration timescale and 
the dynamical timescale of  the shocked pulsar wind. 
In such a case, the cut-off energy of the synchrotron 
radiation is $E<E_{\mathrm{syn}}\sim 200$~MeV, and the 0.1-1~GeV flux 
is sensitive to the magnetic field. 
Furthermore, contributions by emissions from the high-order generated 
pairs \citep{Sierpowska2008} and the 
inverse-Compton emission from the unshocked pulsar wind 
\citep[e.g.,][]{Khangulyan2011a} to the observed spectrum 
below the TeV band have been pointed out. 
Therefore, a more detailed modeling of the 0.1-100~GeV emission process is 
necessary for a closer comparison with the $Fermi$ results. 

 For the TeV emissions, which are produced by the inverse-Compton process, 
 we can see in Figure~\ref{depen} that the second peak of 
 the light curve aligns with
 that of the X-ray band, whereas the first peak comes 
 around near the periastron.  The timing of the first peak 
 of the TeV lightcurve reflects the fact 
 that the stellar soft-photon field strength at 
 the emission region (that is, the shock) reaches its maximum 
 during periastron passage. 
 As we described in section~\ref{x-ray}, 
 though the conversion efficiency  from the spin down energy   
 to the  particles energy of the shocked wind decreases as the 
 pulsar moves toward the periastron, 
 the effect of the increasing strength of the soft photon fields  
toward  the periastron outweighs the decrease of the conversion efficiency. 
 The periastron peak does not well reproduce the observations, 
 at least of the  2004 periastron passage.

It is shown in Figure 5 that the light curve in the $>$ 300 GeV band suffers from photon-photon absorption where soft photons are assumed 
to come only from the Be star.  In the present calculation,  we  have taken 
 a spherically symmetric stellar photon field. 
However,  \citet{Negueruela2011} estimated a 
stellar temperature of $T_{\mathrm{eff}}\sim 3.4\times 10^{4}$~K at the pole 
and $T_{\mathrm{eff}}\sim 2.75\times 10^{4}$~K on the equator, 
which implies that the stellar photon field does depend on the latitude. 
Such a latitudinal dependence of the stellar photon field could affect the 
peak phase of $> 300$~GeV emissions.  
As mentioned in section~\ref{model}, the current model also omits 
the contribution from the disk emission, but it can 
become important for the spectrum and the absorption of the GeV-TeV 
photons. For example,  \citet{vanSoelen2011} studied 
the effects of the IR excess  on the spectrum and found that  
the GeV gamma-ray flux can increase by a factor $\ge 2$. 
 When we include the contribution from the Be-disk emission, 
the absorption effect could be more than doubled, as inferred
 from \citet{vanSoelen2011}. Due to this effect, the flux 
of $>$ 300 GeV gamma-rays around the periastron phase may be
 suppressed. As a result, the peak phase of the 
$>$ 300 GeV gamma-ray light curve may be shifted from the periastron 
phase to a phase prior to the periastron. We are planning to investigate 
this effect using a Monte-Carlo radiative transfer code as our next step.

\subsection{Multi-wavelength Spectra}
\begin{figure}
\begin{center}
\includegraphics[width=15cm,height=7cm]{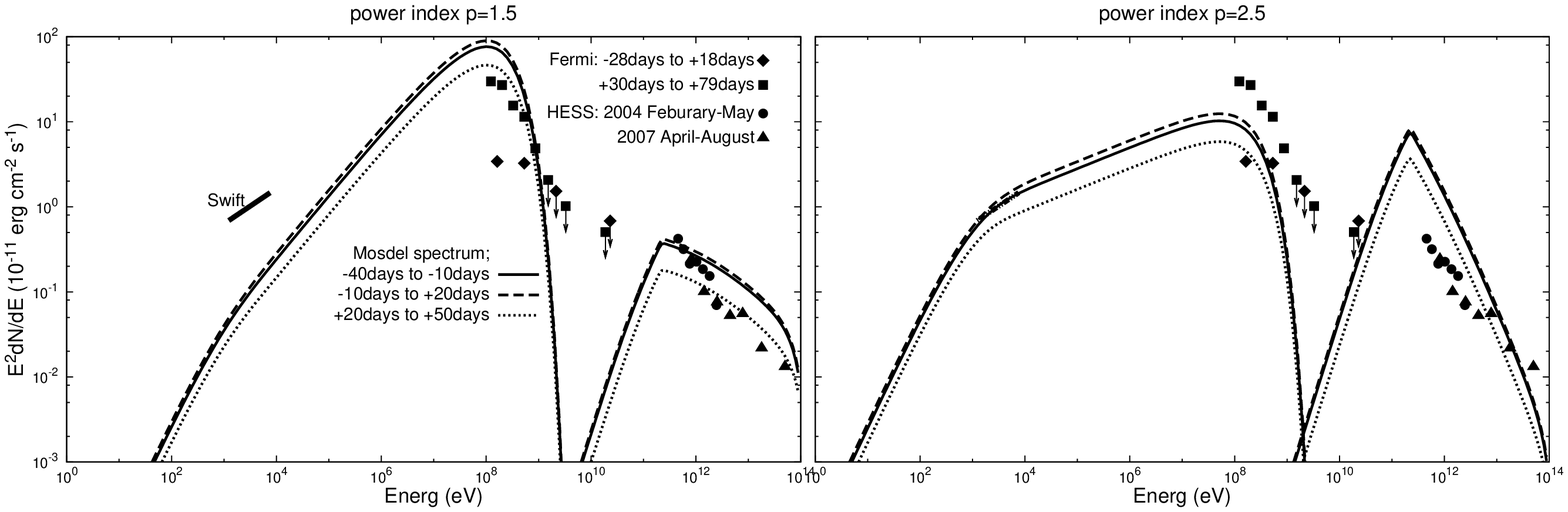}
\caption{Spectrum of the emissions during the periastron passage. The left and 
right panels show the model spectra for  $p=1.5$  and  $p=2.5$, respectively. 
The solid, dashed and dotted lines 
represents the model spectra averaged over  
 three different periods  
$\tau=-40\;\mathrm{d}--10\;\mathrm{d}$, $\tau=-10\;\mathrm{d} - 
+20\;\mathrm{d}$, and $\tau=+20\;\mathrm{d} - +50\;\mathrm{d}$, respectively. 
The data points are taken from Aharonian et al. (2005, 2009) for the 
H.E.S.S. observations between  February and May 2004 and between April 
and August 2007,  and from  Abdo et al. (2011) for the Fermi observations 
and Swift observation of the 2010-2011 periastron passage.}
\label{spectrum}
\end{center}
\end{figure}

 In Figure~\ref{spectrum},   we show 
the  spectra during the periastron passage 
 in multi-wavelength. On top of them, we plot the observed spectra measured 
by various instruments. 
 To account for the systematic uncertainty in our emission model on 
 the shock conditions and hence the energy distribution 
 of the underlying accelerated particles, two model spectra are
 calculated using p=1.5 (left panel) and p=2.5
 (right panel), respectively.  In the figures, the solid, dashed and dotted 
lines represent the spectra averaged over three different period  
$\tau=-40\;\mathrm{d}--10\;\mathrm{d}$, $\tau=-10\;\mathrm{d} - 
+20\;\mathrm{d}$, and $\tau=+20\;\mathrm{d} - +50\;\mathrm{d}$, respectively.
  Because we stopped  our simulation 
at $\tau=+50\;\mathrm{d}$, we cannot calculate the spectrum averaged 
over the period covered by  $Fermi$ observation.  
The model spectra  below and above 1~GeV 
 correspond to the synchrotron emission and the inverse-Compton 
process, respectively.  
In Figure~\ref{spectrum}, we can  see  a typical flux and spectral 
shape for each of different power indexes.  
The spectral slope does not change much from phase 
to phase if the power index is constant, although the flux level varies. 
  Thus having model spectra  will  help us to infer the actual 
power index  of the energy distribution of the shocked wind particles from the 
photon index of observed spectra.

 The present model shows a spectral break of the synchrotron 
 radiation around 1--10~keV, which corresponds to the minimum 
 Lorentz factor of the particles, $\Gamma_{\mathrm{min}}=5\times 10^{5}$. 
 This spectral feature is consistent with the 
 break around 1~keV measured by the SUZAKU observation 
 \citep{Uchiyama2009}. 
As discussed in section~4.3 and also seen in Figure~\ref{spectrum},
 the present model expects that 
the synchrotron spectrum extends up to the maximum photon energy 
of  $E_{\mathrm{syn}}=27m_ec^3h/(16\pi e^2)\sim 200$~MeV, 
which corresponds to the Lorentz factor of the particles,
 $\Gamma_{\mathrm{max}}\sim 
(9m_e^2c^4/4e^3B)^{1/2}=3.6\times 10^{8}(\eta/0.1)^{-1/4}(P_{\mathrm{tot}}/
0.1\mathrm{dyne~cm^{-2}})^{-1/4}$, where we used the equation of 
(\ref{magnetic}).

Although  the flare-like GeV emissions detected by the $Fermi$ telescope 
may not be compared with the result of the 
present simple calculation, where we ignore the 
effects of the radiative cooling.  We would like to remark that  
a smaller power index of the particle distribution 
is preferred to explain the observed flux of $>$100~MeV emissions.
In fact,  Takata \& Taam (2009), who fit the observed X-ray 
flux and the photon index for various orbital phases,   pointed out 
that the expected $>$100~MeV flux for the smaller power index can be 
higher than the $Fermi$ sensitivity.  A more detailed 
modeling for  the spectrum and light curve in GeV energy 
bands will be done in our subsequent studies. 

 In the present calculation, the inverse-Compton process  between 
 the accelerated particles with  $\Gamma_{min}>5\times 10^{5}$ and the 
 stellar photons of energy $\sim 1$~eV takes place 
 in the Klein-Nishina regime.  
 In such a case, the inverse-Compton spectrum calculated with 
 the particle power index $p$ is given by $EF_{E}\propto E^{1-p}$ above 
the energy $\sim \Gamma_\mathrm{min}m_ec^2\sim 2\times 10^{11}~\mathrm{eV}$.  
As we can  see in Figure~\ref{spectrum}, the observed spectrum  
 $EF_E\propto E^{-1}$ by H.E.S.S.  (filled circles and triangles)
  indicates the power  low index $p$ of the distribution of the  scattering
 particles  is $p\sim 2$. The present model also predicts that 
there is a change of the spectral slope at 
 $\sim5\times 10^{11}~\mathrm{eV}$. 

\subsection{Effect of Radiative Cooling}
 In the shocked regions with strong magnetic fields, 
the synchrotron cooling timescale $t_s$ of the accelearated particles may  
 be less than the crossing time $t_c$ of those regions.
In the present framework, using equations~(\ref{magnetic}),  
the total pressure and hence the magnetic field strength 
become largest when the pulsar
penetrates the disk, so that the effect of synchrotron cooling 
may not be neglected at that phase. In addition, 
the inverse-Compton cooling time may be less than $t_c$ at 
phases near the periastron. These cooling processes  can 
affect the resulting spectrum and light curve. For example,  
if the inverse-Compton process dominates the other cooling processes 
of the TeV particles,  
the X-ray emission via synchrotron radiation is  weaker in 
denser soft photon fields. On the other hand, if the synchrotron  
process is the dominant cooling process of the TeV particles,  
the TeV emissions via the inverse-Compton process is weaker 
in regions with stronger magnetic fields.  

For PSR~B1259-63/LS~2883, however,
 we expect that the cooling processes  have no important effect 
on the X-ray light curve.
   The synchrotron radiation occurs
  in the slow cooling regime 
 if the particle's  Lorentz factor
 is smaller than $\Gamma_\mathrm{sy}=(9m_e^3c^6)/(4e^4B^2 \ell)$, 
where $\ell$ is the size of the emission cavity. With 
 $B\sim 0.1~$Gauss and $\ell\sim 10^{13}$~cm as 
 typical values near the periastron,  the critical  
  energy below which the synchrotron cooling process can be ignored 
is $E\sim 7.62\times 10^{7}(B/0.1\mathrm{G})^{-3}(\ell/10^{13}
\mathrm{cm})^{-2})$~eV, which is  way above the X-ray energy band. 
Moreover, according to \citet{Tavani1997}, 
the inverse-Compton process  between the
accelerated particles and stellar photons enhances the double-peaked structure 
in the X-ray light curve. Hence, our model X-ray light curves remain 
robust even if the detailed cooling processes will be taken into account.

\section{Summary}
\label{summary}
In our previous study \citep{Okazaki2011},
we have developed 3-D SPH simulations  of the interaction 
between the pulsar wind and the Be-disk and wind in the gamma-ray binary 
PSR~B1259-63/LS~2883.  In this paper, 
we investigated the high-energy emissions from the
shocked pulsar wind,
calculating the synchrotron radiation and the 
inverse-Compton process 
on the basis of the simulated shock geometry and pressure distribution 
of the pulsar wind.   The current study revealed that
the observed double-peaked X-ray light curves are reproduced only if 
the Be-disk is denser than typical (with base density 
$\sim 10^{-9}\,\mathrm{g~cm^{-3}}$).
The pre- and post-periastron X-ray peaks appear respectively
when the pulsar passes through the disk 
prior to the periastron, and when the pulsar wind creates a cavity 
in the disk gas after the periastron, in both cases 
terminating the pulsar wind over 
a large solid angle around the pulsar.

 On the other hand, in the model TeV light curve, which also shows a 
double peak feature, the first peak appears around the periastron,
 which will disagree with the 2004 H.E.S.S. observation  showing  
the first peak located at a pre-periastron phase.
In a subsequent paper, we will study whether 
the effects of the disk emission to the inverse-Compton process and
photon-photon absorption process can shift the first peak to a phase 
prior to the periastron passage. 

\bigskip
We express our appreciation to an anonymous referee for  useful  comments.
 J.T. thanks K.S. Cheng and  R.E. Taam  for useful discussions. 
S.N. is supported by Grant-in-Aid for Scientific Research on Innovative
Areas No. 23105709 by Ministry of Education, Culture, Sports,
Science and Technology (MEXT), Grant-in-Aid for Scientific Research (S)
No. 19104006 and Scientific Research (B) No. 23340069
by Japan Society for the Promotion of Science (JSPS),
Joint Usage/Research Center for Interdisciplinary Large-scale
  Information Infrastructures in Japan 
and Grant-in-Aid for the Global COE Program 
"The Next Generation of Physics, Spun from Universality and Emergence" 
from MEXT of Japan. T.N. is supported by Grant-in-Aid for Scientific 
Research (C) No. 23540271 by Japan Society for the Promotion of Science (JSPS).
S.P.O acknowledges partial support from grant \#NNX11AC40G from NASA's 
Astrophysics Theory Program. The computation was carried
out on HITACHI SR16000 at Yukawa Institute for Theoretical Physics (YITP),
Kyoto University and on HITACHI SR11000 at the Information 
Initiative Center (iiC), Hokkaido university. 
 In addition to the above grants, this 
work was partially supported by 
the iiC collaborative research program 2010-2011, the Grant-in-Aid for
Scientific Research (18104003, 19047004, 19740100, 20540236, 21105509,
21540304, 22340045, 22540243, 23105709), and a research grant from
Hokkai-Gakuen Educational Foundation.

\clearpage



\newpage

\begin{thebibliography}{}
\bibitem[Abdo et al.(2011)]{Abdo2011} Abdo et al. 2011, \apjl,  736, 11
\bibitem[Aharonian et al.(2005)]{Aharonian2005}
   Aharonian, F. et al. 2005, \aap, 442, 1 
\bibitem[Aharonian et al.(2006)]{Aharonian2006} 
   Aharonian, F. et al. 2006, \aap, 460, 743
\bibitem[Aharonian et al.(2009)]{Aharonian2009} 
   Aharonian, F. et al. 2009, \aap, 507, 389
\bibitem[Albert et al.(2006)]{Albert2006} Albert, J. et al. 2006, Sci, 312, 1771
\bibitem[Bate et al.(1995)]{Bate1995}
   Bate M.R., Bonnell I.A., Price N.M., 1995, \mnras, 285, 33
\bibitem[Begelman \& Sikora (1987)]{Bege1987}
Begelman, M.C. \& Sikora, M., 1987, ApJ, 322, 650

\bibitem[Bogovalov et al.(2011)]{Bogovalov2011} 
Bogovalov, S.V., Khangulyan, D., Koldoba, A.V., Ustyugova, G.V. 
\& Aharonian, F. A., 2011, MNRAS, in press
\bibitem[Bogovalov et al.(2008)]{Bogovalov2008} 
   Bogovalov, S.V., Khangulyan, D.V., 
   Koldoba, A.V., Ustyugova, G.V. \&  Aharonian, F.A. 2008, \mnras, 387, 63
\bibitem[Bongiorno et al.(2011)]{Bongiorno2011} 
   Bongiorno, S.,  Falcone, A.,  Stroh, M.,  Holder, J.,  Skilton, J.,
   Hinton, J.,  Gehrels, N., \&  Grube, J., 2011, arXiv:1104.4519B
\bibitem[Carciofi \& Bjorkman(2006)]{Carciofi2006}
   Carciofi, A.C., \& Bjorkman, J.E., 2006, \apj, 639, 1081
\bibitem[Chernyakova et al.(2006)]{chernyakova06} Chernyakova, M., Neronov, A., Lutovinov, A.,
Rodriguez, J., Johnston, S., 2006, \mnras, 367, 1201
\bibitem[Connors et al.(2002)]{connors02} Connors, M., Chodas, P., Mikkola, S.,
Manchester, R.N., McConnell, D., 2002, \mnras, 37, 1435
\bibitem[Corbet et al.(2011)]{Corbet2011} Corbet R.H.D. et al.. 2011,  ATel, 3221, 1
\bibitem[Dubus(2006)]{Dubus2006} Dubus, G. 2006, \aap, 451, 9
\bibitem[Dubus(2010)]{Dubus2010a}
   Dubus, G. 2010, High Energy Phenomena in Massive Stars ASP Conference Series, 
   Vol. 422, proceedings of a conference held 2-5 February 2009 
\bibitem[Dubus et al.(2010)]{Dubus2010b} 
   Dubus, G., Cerutti, B., Henri, G., 2010,  \aap, 516 18
\bibitem[Ferland(1996)]{Ferland1996}
   Ferland G.J., 1996, CLOUDY: 90.01
\bibitem[Hinton et al.(2009)]{Hinton2009}
   Hinton et al. 2009, \apjl, 690L, 101
\bibitem[Hirayama et al.(1996)]{Hirayama1996} 
   Hirayama, M., Nagase, F., Tavani, M., 
    Kaspi, V.M., Kawai, N., \&  Arons, J. 1996, \pasj, 48, 833
\bibitem[Johnston et al.(1994)]{Johnston1994}
   Johnston S., Manchester R.N., Lyne A., Nicastro, L., 
   Spyromilio, J., 1994, \mnras, 268, 430
\bibitem[Johnston et al.(2005)]{Johnston2005}
   Johnston, S., Ball, L., Wang, N., 
   \& Manchester, R.N. 2005, \mnras, 358, 1069
\bibitem[Kawachi et al.(2004)]{Kawachi2004} Kawachi A. et al.,  2004, ApJ, 607, 949
\bibitem[Kennel \& Coroniti(1984)]{Kennel1984}
   Kennel, C.F., \& Coroniti, F.V. 1984, \apj, 283, 694
\bibitem[Khangulyan et al.(2007)]{Khangulyan2007}
   Khangulyan, D., Hnatic, S., Aharonian, F., Bogovalov, S., 2007, \mnras, 380, 320
\bibitem[Khangulyan et al.(2011a)]{Khangulyan2011a}
   Khangulyan, D., Aharonian, F.,  Bogovalov, S.,  Rib\'{o}, M., 
   2011, arXiv:1104.0211
\bibitem[Khangulyan et al.(2011b)]{Khangulyan2011b}
   Khangulyan, D., Hnatic, S., Aharonian, F., 
   \&  Bogovalov, S.  2007, \mnras, 380, 320
\bibitem[Kong et al.(2011)]{Kong2011}
   Kong, S.W., Yu, Y.W., Huang, Y.F. \& Cheng, K. S., 2011, 
   arXiv:1105.3900
\bibitem[Lee et al.(1991)]{Lee1991} 
   Lee, U., Saio, H., Osaki, Y., 1991, \mnras, 250, 432
\bibitem[Longair (1994)]{Longair1994} 
   Longair, M. S. 1994, in High Energy Astrophysics, Vol. 2 (2nd ed.; Cambridge:
Cambridge Univ. Press), 357:section 21.5
\bibitem[Mold\'{o}n et al.(2011)]{Moldon2011}
   Mold\'{o}n, J., Johnston, S.,  Rib\'{o}, M.,  Paredes, J.M., Deller, A.T., 2011, \apjl, 732, 10
\bibitem[Nagataki(2004)]{Nagataki2004} Nagataki, S., 2004, \apj, 600, 883
\bibitem[Negueruela et al.(2011)]{Negueruela2011}
   Negueruela, Ignacio; Rib\'{o}, M.,  Herrero, A., Lorenzo, J.,  Khangulyan, D., Aharonian, F.A., 
   2011, \apjl, 732, 11
\bibitem[Neronov \& Chernyakova(2007)]{Neronov2007}
   Neronov, A., Chernyakova, M., 2007, \apss, 309, 253
\bibitem[Okazaki et al.(2002)]{Okazaki2002} 
   Okazaki, A.T., Bate, M.R., Ogilvie, G.I. \&  Pringle, J. E., 
   2002, \mnras, 337, 967O
\bibitem[Okazaki et al.(2011)]{Okazaki2011} 
   Okazaki, A. T., Nagataki, S., Naito, T., Kawachi, A., Hayasaki, K.,
   Owocki, S. P., Takata, J. 2011, \pasj, 63, 893 (paper I)
\bibitem[Okazaki et al.(2011)]{Okazaki2012} 
Okazaki, A. T. et al. 2012, in prepare 
\bibitem[Porter(1999)]{Porter1999}
   Porter J. M. 1999, \aap, 348, 512
\bibitem[Sari et al. (1998)]{sari98} Sari, R., Piran, T., Narayan, R. 1998 ApJL 497 L17
\bibitem[Sierpowska-Bartosik \&  Bednarek(2008)]{Sierpowska2008}
   Sierpowska-Bartosik, A. \&  Bednarek, W. 2008, \mnras, 385, 2279
\bibitem[Silaj et al.(2010)]{Silaj2010}
   Silaj, J., Jones, C. E., Tycner, C., Sigut, T. A. A., Smith, A. D., 2010,
   \apjs, 187, 228
\bibitem[Uchiyama et al.(2009)]{Uchiyama2009}	Uchiyama, Y.,  Tanaka, T., 
   Takahashi, T.,  Mori, K., \&  Nakazawa, K. 2009 \apj, 698, 911
\bibitem[Takata \& Taam(2009)]{Takata2009} Takata, J., Taam, R.E., 2009, \apj, 702,100
\bibitem[Tam et al.(2011)]{Tam2011}
   Tam, P.H.T.,  Huang, R.H.H.,  Takata, J.,  Hui, C.Y.,  Kong, A.K.H.,
    Cheng, K. S., 2011, \apjl, 736, 10
\bibitem[Tavani \& Arons(1997)]{Tavani1997}
   Tavani, M., \& Arons, J. 1997, \apj, 477, 439
\bibitem[van Soelen \&  Meintjes(2011)]{vanSoelen2011}
   van Soelen, B., \&  Meintjes, P.J., 2011, \mnras, 412, 1721
\bibitem[Xu et al. (2011)]{xu11} Xu, M., Nagataki, S., Huang, Y.F., 2011, ApJ 735:3

\end{thebibliography}
\end{document}